\documentclass[%
	10pt,%
	amsmath,%
	amsfonts,%
	amssymb,%
	aps,
	pra,%
	twocolumn,%
	superscriptaddress,
]{revtex4-2}
\usepackage[utf8]{inputenc}
\usepackage[T1]{fontenc}
\usepackage[english]{babel}
\usepackage{csquotes}
\usepackage{lmodern}
\usepackage{wrapfig}

\usepackage{xr}
\usepackage[%
  colorlinks=true,%
	citecolor=blue,%
	linkcolor=black,%
	urlcolor=blue%
]{hyperref}

\usepackage{graphicx}
\usepackage{siunitx} 
\graphicspath{{Figures/}}

\usepackage{bm}
\usepackage{upgreek}

\def\ifnonemptyparenthesis#1{%
  \if\relax\detokenize{#1}\relax%
  \else%
    (#1)%
  \fi%
}

\AtBeginDocument{}
\AtBeginDocument{}

\newcommand{\fig}[2][]{\autoref{#2}\ifnonemptyparenthesis{#1}}

\begin{document}

\title{Fast long-distance transport of cold cesium atoms}

\author{Till~Klostermann}
\author{Cesar~R.~Cabrera}
\author{Hendrik~von Raven}
\author{Julian~F.~Wienand}
\affiliation{Fakult\"at f\"ur Physik, 
             Ludwig-Maximilians-Universit\"at, 
             Schellingstr.\ 4, D-80799 M\"unchen, Germany}
\affiliation{Munich Center for Quantum Science and Technology (MCQST),
             Schellingstr. 4, D-80799 M\"unchen, Germany}
\affiliation{Max-Planck-Institut f\"ur Quantenoptik, 
             Hans-Kopfermann-Strasse 1, D-85748 Garching, Germany}
\author{Christian~Schweizer}
\affiliation{Fakult\"at f\"ur Physik, 
             Ludwig-Maximilians-Universit\"at, 
             Schellingstr.\ 4, D-80799 M\"unchen, Germany}
\affiliation{Munich Center for Quantum Science and Technology (MCQST),
             Schellingstr. 4, D-80799 M\"unchen, Germany}
\affiliation{Walther-Meißner-Institut, Bayerische Akademie der Wissenschaften, D-85748 Garching, Germany}
\author{Immanuel~Bloch}
\affiliation{Fakult\"at f\"ur Physik, 
             Ludwig-Maximilians-Universit\"at, 
             Schellingstr.\ 4, D-80799 M\"unchen, Germany}
\affiliation{Munich Center for Quantum Science and Technology (MCQST),
             Schellingstr. 4, D-80799 M\"unchen, Germany}
\affiliation{Max-Planck-Institut f\"ur Quantenoptik, 
             Hans-Kopfermann-Strasse 1, D-85748 Garching, Germany}
\author{Monika~Aidelsburger}
\affiliation{Fakult\"at f\"ur Physik, 
             Ludwig-Maximilians-Universit\"at, 
             Schellingstr.\ 4, D-80799 M\"unchen, Germany}
\affiliation{Munich Center for Quantum Science and Technology (MCQST),
             Schellingstr. 4, D-80799 M\"unchen, Germany}

\begin{abstract}
Transporting cold atoms between distant sections of a vacuum system is a central ingredient in many quantum simulation experiments, in particular in setups, where a large optical access and precise control over magnetic fields is needed. In this work, we demonstrate optical transport of cold cesium atoms over a total transfer distance of about \SI{43}{cm} in less than $\SI{30}{ms}$. The high speed is facilitated by a moving lattice, which is generated via the interference of a Bessel and a Gaussian laser beam. We transport about $3\times 10^6$ atoms at a temperature of a few~\si{\micro K} with a transport efficiency of about $\SI{75}{\%}$. We provide a detailed study of the transport efficiency for different accelerations and lattice depths and find that the transport efficiency is mainly limited by the potential depth along the direction of gravity. To highlight the suitability of the optical-transport setup for quantum simulation experiments, we demonstrate the generation of a pure Bose-Einstein condensate with about $2\times 10^4$ atoms. We find a robust final atom number within $2\%$ over a duration of \SI{2.5}{h} with a standard deviation of $<5\%$ between individual experimental realizations. 
\end{abstract}

\date{\today}

\maketitle

\section{Introduction}

Ultracold atoms in optical lattices are powerful platforms for quantum simulation of complex quantum many-body systems~\cite{georgescu_quantum_2014}, most notably in the context of condensed matter physics~\cite{grossQuantumSimulationsUltracold2017,schaferToolsQuantumSimulation2020,browaeys_many-body_2020}. Recent advances further opened up promising new directions~\cite{schweizer_floquet_2019,surace_lattice_2020,mil_scalable_2020,yang_observation_2020,periwal_programmable_2021} to study phenomena related to quantum gravity~\cite{bentsen_treelike_2019,belyansky_minimal_2020,brown_quantum_2021}, quantum electrodynamics and high-energy physics~\cite{banuls_simulating_2020,aidelsburger_cold_2021}. 
The precise parameter control and natural scalability of neutral-atom devices enables analog quantum simulation with hundreds of atoms, which significantly challenges the limits of state-of-the-art numerical methods~\cite{trotzky_probing_2012,ebadi_quantum_2021,scholl_quantum_2021,hebbe_madhusudhana_benchmarking_2021}. 
Moreover, neutral atoms offer a broad range of applications in quantum metrology and sensing~\cite{ludlow_optical_2015} and digital quantum computing~\cite{henriet_quantum_2020}.  

\begin{figure}[t!]
\includegraphics{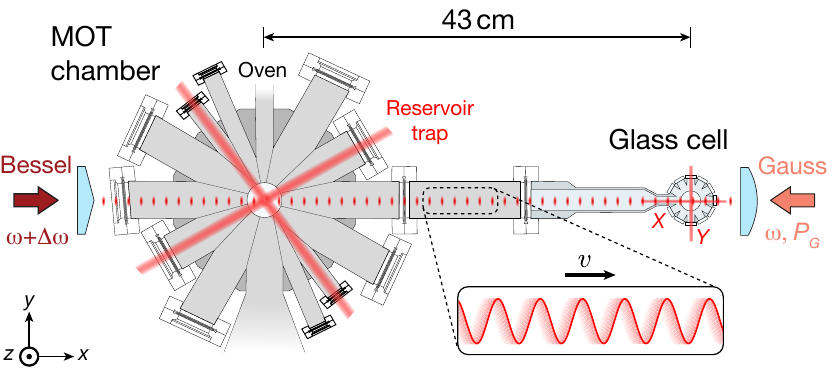}
\caption{\textbf{Experimental setup}. Vacuum system with two-chamber design: pre-cooling in a magneto-optical trap (MOT) is performed in the first chamber, the final evaporation to Bose-Einstein condensation takes place in the glass cell. The distance between both chambers is $43\,$cm. The atom source/oven is connected to the top port.
The crossed-dipole beams, which form the reservoir trap in the MOT chamber and the dipole trap ($X$, $Y$) for evaporation in the glass cell, are shown in red. The transport lattice (red dots) is formed by interfering two counter-propagating laser beams: a Bessel beam (dark red arrow) with frequency $\omega + \Delta \omega$ (left), which is generated by sending a laser beam with $35\,$W onto an axicon and a Gaussian laser beam (light red arrow) with frequency $\omega$ and variable power $P_G$, which is focused between the two chambers by a lens.
The detuning $\Delta\omega = 2\pi\Delta f$ results in a running-wave lattice, that moves with velocity $v$.}%
\label{fig:1}%
\end{figure}

For high-fidelity operation large optical access~\cite{gross_quantum_2020}, a clean electromagnetic environment and excellent vacuum conditions are indispensable. This is often achieved by separating the main experimental apparatus into two distinct sections: a first chamber for preparation and pre-cooling of the cold atomic cloud, and a second science chamber where the actual measurements are performed (\fig{fig:1}). 
Such a two-chamber design naturally requires transporting the atoms between the two separate vacuum sections. 
However, this usually comes at the expense of increased experimental complexity and longer cycle times. On the other hand reaching faster cycle times~\cite{kinoshita_all-optical_2005,stellmer_laser_2013,stellmer_production_2013,roy_rapid_2016,hu_creation_2017,urvoy_direct_2019,solano_strongly_2019,phelps_sub-second_2020} and developing compact and robust experimental setups~\cite{lam_compact_2020} is essential for the development of the next generation of quantum devices~\cite{komar_quantum_2014,koller_transportable_2017,bongs_taking_2019,altman_quantum_2019}. 

There are various different transport schemes that have been developed, which either make use of magnetic~\cite{greinerMagneticTransportTrapped2001,lewandowski_observation_2002,pertotVersatileTransporterApparatus2009}, optical~\cite{schraderOpticalConveyorBelt2001,schmidLongDistanceTransport2006a,marchantGuidedTransportUltracold2011,middelmannLongrangeTransportUltracold2012,leonardOpticalTransportManipulation2014,grossAllopticalProductionTransport2016,langbeckerHighlyControlledOptical2018,unnikrishnan_long_2021} or hybrid~\cite{marchantGuidedTransportUltracold2011,middelmannLongrangeTransportUltracold2012} traps. 
Magnetic transport typically relies on the translation of the trap minimum either by dynamically controlling the current in overlapping pairs of coils~\cite{greinerMagneticTransportTrapped2001} or by mechanically moving a single pair of coils~\cite{lewandowski_observation_2002,pertotVersatileTransporterApparatus2009}. 
While magnetic transport has been demonstrated reliably for large distances, it requires complex mechanical engineering, typically limits the optical access, and is only applicable to magnetically trappable atoms.
Optical transport, on the other hand, can be implemented for any atomic species, although typically at reduced trap depths. 
The most straightforward implementation is based on a mechanically-movable lens, that generates a tightly-focused optical dipole trap with variable focus position~\cite{grossAllopticalProductionTransport2016,couvertOptimalTransportUltracold2008}. 
Since moving mechanical parts introduce vibrations, novel schemes based on focus-tunable lenses~\cite{leonardOpticalTransportManipulation2014,unnikrishnan_long_2021} have been developed. 
However, the total transport duration for all schemes mentioned above is fundamentally limited either by small longitudinal trapping frequencies or by the finite velocities of the mechanical stages, which results in transport times on the order of a second for typical transport distances. 
This motivates the use of running-wave optical lattices, where the motion is controlled via the frequency detuning of two counter-propagating laser beams. 
This configuration offers large longitudinal trapping frequencies and no moving components are needed~\cite{schraderOpticalConveyorBelt2001,schmidLongDistanceTransport2006a,langbeckerHighlyControlledOptical2018}. 

Here we report fast optical transport of $^{133}$Cs atoms over an unprecedented distance of \SI{43}{cm} (\SI{86}{cm} roundtrip) in <$\,30\,$ms ($60\,$ms) using a far-detuned running-wave optical lattice, where we reach final velocities of up to $26.6\,$m/s. 
Due to the large mass of cesium, large optical gradients are required during transport. 
Therefore, the running-wave lattice is generated by interfering a Bessel and a Gaussian laser beam (\fig{fig:1}), similar to Ref.~\cite{schmidLongDistanceTransport2006a}.  
We observe a one-way transport efficiency of $\sim$75$\%$ and demonstrate the robustness of the scheme by generating a Bose-Einstein condensate (BEC). The mean atom number is stable within $2\%$ over the course of \SI{2.5}{h}, making this scheme suitable for state-of-the-art quantum simulation experiments using heavy atoms such as $^{133}$Cs.

\section{Experimental Setup}
Our experimental apparatus consists of two main vacuum chambers separated by \SI{43}{cm} (Fig.~\ref{fig:1}). The MOT chamber is a steel chamber, where all pre-cooling steps are performed: MOT, optical molasses and degenerate Raman-sideband cooling~(Sec.~\ref{sec:sequence} and Fig.~\ref{fig:S6} in~\cite{supplement}). The science chamber is a glass cell with large optical access in order to support single-atom single-site manipulation and read-out~\cite{bakr_quantum_2009, shersonSingleatomresolvedFluorescenceImaging2010, yamamotoYtterbiumQuantumGas2016, cheukQuantumGasMicroscopeFermionic2015, hallerSingleatomImagingFermions2015, edgeImagingAddressingIndividual2015, parsonsSiteResolvedImagingFermionic2015, omranMicroscopicObservationPauli2015,alberti2016} using a high numerical aperture (NA) objective that is placed outside the vacuum chamber. Moreover, it has 11 side ports, which can be used for optical lattices and additional dipole potentials.

\paragraph*{Optical transport setup}
The large mass $m$ of $^{133}$Cs requires large optical gradients to hold the atoms against gravity. 
The diffraction of Gaussian beams makes this challenging to achieve because large laser powers are required for a sufficiently steep trap over the full transport distance.
To circumvent this issue we employ a Bessel beam that is generated using an axicon~\cite{schmidLongDistanceTransport2006a}. 
It has a diffractionless range $x_B = w_0\tan(\alpha/2)/(n-1)$, which depends on the apex angle $\alpha$ of the axicon, the waist $w_0$ of the incoming Gaussian laser beam and the axicon's refractive index $n$ [\fig[a]{fig:2}], as described e.g. in Refs~\cite{schmidLongDistanceTransport2006a,brzobohatyHighQualityQuasiBessel2008}. 
Within the range $x_B$ the radius of the central spot of the beam profile remains approximately constant [\fig[a]{fig:2} and Fig.~\ref{fig:besselradius} in~\cite{supplement}]. 
For an ideal axicon, the $1/e^2$ waist of the central peak is $w_B\simeq 1.8\cos(\alpha/2)/k(n-1)$ and its intensity follows a Gaussian shape; $k=2\pi / \lambda$ is the wavevector and $\lambda$ is the wavelength of the laser beam. 
Imperfections at the tip of the axicon result in additional oscillations of the intensity along the transport axis [\fig[b]{fig:2} and Fig.~\ref{fig:besselsim} in~\cite{supplement}].
In our case we use an axicon with an apex angle $\alpha=179$\SI{}{\degree} and a laser beam with $\lambda=1064$\SI{}{nm} and $w_0=2.5\,$mm. This results in a central spot with waist $w_B=\SI{80}{\micro m}$ and sufficient peak intensity over the full transport distance of 43\,cm [\fig[b]{fig:besselradius} in~\cite{supplement}]. 
The Bessel beam acts as a waveguide that holds the atoms against gravity and the interference with the Gaussian laser beam forms the transport lattice. 
For the latter we employ a rather large waist of \SI{600}{\micro m} and find that a moderate power of $P^{\text{typ}}_G=6.5$\SI{}{W} is enough for optimal transport efficiencies. 
The focus position of the beam was optimized by maximizing the transport efficiency. The optimum was found about \SI{21}{cm} away from the center of the glass cell.

\paragraph*{Control of the detuning}

The two counter-propagating laser beams form a one-dimensional (1D) optical lattice $V(x)\propto \sqrt{P_G} \sin(2kx + 2\pi \Delta f\,t/2)$.
Here $\Delta f$ is the frequency difference between the two beams, which results in a moving lattice with velocity $v= \lambda \Delta f/2$. 
For fast and accurate transport, the detuning needs to be accurately controlled with a large bandwidth.
This is achieved using direct digital synthesis (DDS) and AOMs (acousto-optic modulator) in double pass configuration (DPAOMs) as illustrated in \fig[c]{fig:2}.

In order to avoid a frequency offset lock, we use two separate fiber amplifiers at \SI{1064}{nm} to generate the transport laser beams. We split off light from one of them and use it to seed the second amplifier.
In between, we implement a frequency-shifting setup that consists of two DPAOMs with dynamically tunable frequencies $f_1$ and $f_2$. 
This enables broadband tuning, where each DPAOM can introduce a detuning up to $\Delta f=\SI{50}{MHz}$, which corresponds to a lattice velocity $v \approx \SI{26.6}{m/s}$.
In combination with the large trap frequencies in the lattice along the transport direction, accelerations up to several thousand \si{m/s^2} can be achieved.

\begin{figure}[t!]
\includegraphics{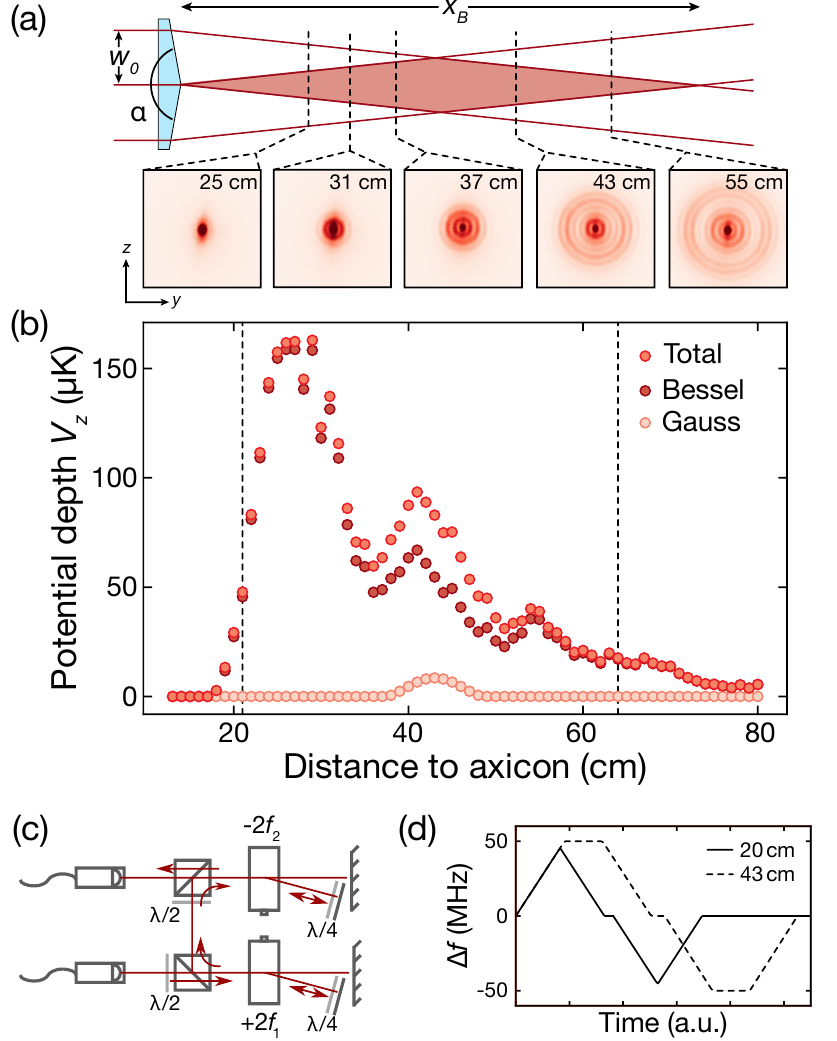}
\caption{\textbf{Transport setup}. (a) Realization of the Bessel beam with an axicon with apex angle $\alpha$ and an incoming Gaussian laser beam with waist $w_0$.  
2D images show measured beam profiles at different distances from the axicon. The diffractionless range is given by $x_B$.
(b) Vertical trap depth $V_z$ of the Gaussian transport beam (light red), the Bessel beam (dark red) and the combined trap (red) versus distance from the axicon taking gravity into account. 
The depth is computed using the measured beam profiles of the Bessel beam (power of incoming laser beam: $35\,$W) and the focal position of the Gaussian beam at $P^{\text{typ}}_G=6.5\,$W. 
The dashed vertical lines indicate the start and end point of the transport. 
(c) Schematic setup used to control the frequency offset $\Delta f$ between the two transport beams.
A small amount of light is split off from the laser generating the Gaussian transport beam. Two DPAOM setups with dynamically tunable RF frequencies $f_1(t)$ and $f_2(t)$ are used to introduce frequency detunings up to $50\,$MHz each. The setup is fiber coupled and used to seed a second amplifier, which generates light for the Bessel-shaped transport beam. 
(d) Simplified schematic illustrating the frequency ramps for the roundtrip transport for two different transport distances. 
Positive $\Delta f$ corresponds to transport in the direction of the glass cell.
}%
\label{fig:2}%
\end{figure}

\section{Realization of large-distance optical transport}

\paragraph*{Loading of the transport lattice}
To facilitate loading of low-temperature atoms into the transport lattice, we transfer the dilute, pre-cooled cloud into a large volume reservoir trap and add the Bessel beam at low intensity to act as a tightly-focused optical dimple~\cite{weberBoseEinsteinCondensationCesium2003}. 
The thermalization between the atoms in the reservoir and the dimple (Bessel) enables high densities at low temperatures in the transport lattice. 

The sequence starts by collecting Zeeman-slowed $^{133}$Cs atoms in a MOT within 3$\,$s, which are further cooled in an optical molasses using standard techniques~\cite{supplement}. To further cool and spin-polarize the atoms we use degenerate Raman sideband cooling in a near-detuned optical lattice~\cite{kermanOpticalMolasses3D2000}, which results in a cold atomic cloud of about \SI{2e7}{atoms} in the absolute ground state $|F=3, m_F=3\rangle$ at $<\SI{1}{\micro K}$ with a peak density of $n_0=\SI{3e10}{atoms/cm^3}$.
When loading the atoms adiabatically into a crossed-beam dipole trap, the phase space density (PSD) $\phi=n_0\lambda_{dB}^3$ is conserved; here $\lambda_{dB}=h/\sqrt{2\pi m k_BT}$ denotes the thermal de-Broglie wavelength, $h$ the Planck's constant, $k_B$ the Boltzmann constant and $T$ the temperature of the atoms. 
To limit the temperature increase, the Raman-cooled cloud is transferred into a shallow, large volume dipole trap, which is formed by two crossed laser beams at $1064\,$nm with a circular waist of \SI{0.5}{mm} and a maximum power of \SI{15}{W} and \SI{10}{W}, respectively.
In addition, we apply a magnetic field gradient of \SI{31.3}{G/cm} to hold the atoms against gravity. For efficient thermalization during loading, we further apply a magnetic offset field of \SI{120}{G} to increase the $s$-wave scattering length to $a\approx 1500\,a_0$~\cite{chin_precision_2004}; here $a_0$ denotes the Bohr radius.
After a hold time of \SI{250}{ms} we obtain \SI{6e6}{atoms} at \SI{3}{\micro K} in the crossed dipole trap.


During the next \SI{400}{ms} we ramp up the power of the Bessel beam to \SI{20}{W} and let the atoms re-thermalize.
To load the atoms into the 1D transport lattice, we then redirect all power from the reservoir beams to the Bessel beam within \SI{500}{ms} and
simultaneously ramp up the Gaussian beam to \SI{6.5}{W} to avoid spreading of the cloud along the Bessel beam. The magnetic offset field is changed to $23\,$G to reduce three-body losses.
After the reservoir has been fully removed, we end with about \SI{4e6}{atoms} at \SI{10}{\micro K} in the static transport lattice.
We attribute the large increase in temperature compared to the reservoir trap to compression in the 1D lattice.

\begin{figure}[t!]
\includegraphics{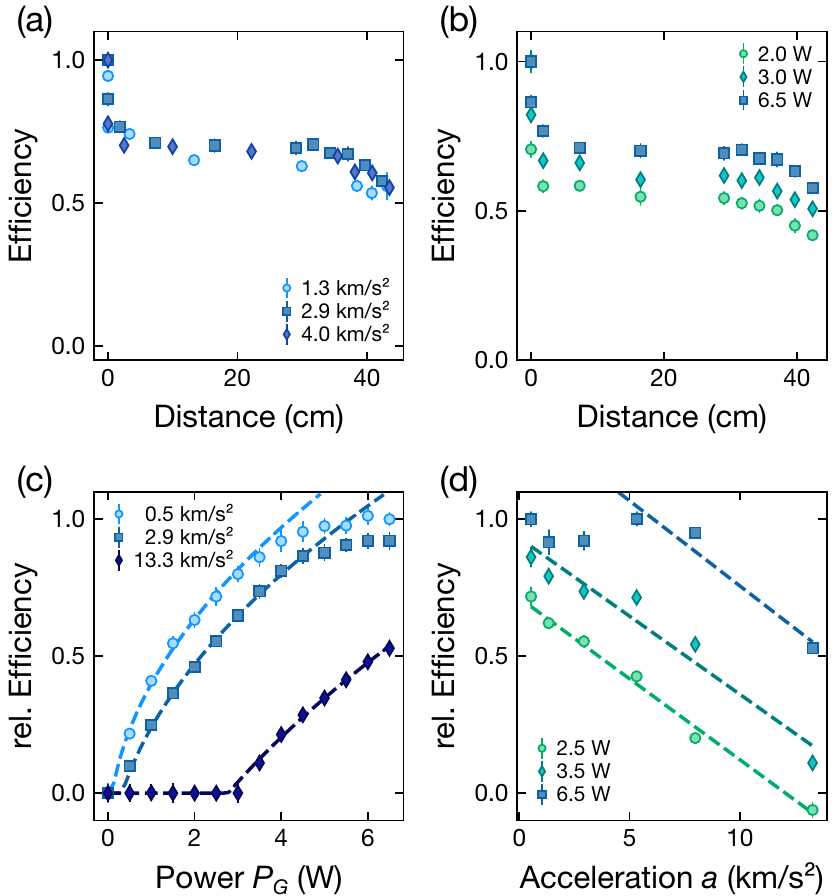}
\caption{\textbf{Transport efficiency}. (a)-(b) Efficiency of roundtrip transport versus transport distance: (a) different accelerations $a$ for $P^{\text{typ}}_G=6.5\,$W and (b) different lattice depths for $a_{\text{typ}}=2.9\,$km/s$^2$. The lattice depth is 
varied by changing the power of the Gaussian transport beam $P_G$.
(c)--(d) One-way transport to the glass cell versus (c) power $P_G$ for three different accelerations $a$ and (d) acceleration $a$ for three different Gaussian powers $P_G$. The dashed lines are guides to the eye.
In (c) the dashed line shows a scaling with $\sqrt{P_G}$ in (d) the scaling is linear with acceleration $a$.
Error bars show the standard error of the mean, extracted from five repetitions in (a)-(d).}%
\label{fig:3}%
\end{figure}

\paragraph*{Transport efficiency}


Optical transport in the running-wave lattice is realized by linearly increasing the detuning $\Delta f$ between the two counter-propagating lattice beams, which results in a constant acceleration $a$.
Before the atoms reach the final position in the glass cell, we apply a linear deceleration ramp, where the detuning is decreased to zero at the same rate as during the acceleration. 
For typical parameters ($a_{\text{typ}}=\SI{2.9}{km/s^2}$, $P^{\text{typ}}_{G}=6.5\,$W), we obtain a final transport velocity of $26.6\,$m/s and the transport duration for the full distance to the glass cell is \SI{25.5}{ms}. 
After transport, we find a total atom number of \SI{3e6}{atoms} in the glass cell at a temperature of \SI{5}{\micro K}. 
This corresponds to a transport efficiency of $\sim$\SI{75}{\%}.
We attribute the reduced temperature in the glass cell to evaporation during transport. The magnetic offset field in the glass cell is set to \SI{28.2}{G} during transport.
Note that there are no additional bias fields along the transport axis. Nonetheless, we do not observe any spin-depolarization.

To investigate the transport efficiency as a function of the transport distance, we perform roundtrip measurements, where the atoms are transported back into the MOT chamber [\fig[a-b]{fig:3}]. 
We scan the final position between the MOT chamber and the glass cell by changing the time between the acceleration and deceleration ramp [\fig[d]{fig:2}].  
At the beginning of the transport (\SI{100}{\micro s}), we observe a sharp decrease of the atom number. Varying the acceleration [\fig[a]{fig:3}] up to \SI{4.0}{km/s^2} does not lead to a significant further change. 
Reducing the lattice depth on the other hand [\fig[b]{fig:3}] results in a significant increase of the atom loss. 
We attribute this initial loss to the sudden onset of the acceleration, which limits the number of atoms that are transported. 
In addition, we observe a reduction of the transport efficiency for the longest distances, which becomes more pronounced for larger accelerations and weaker lattice depths.
This is most likely caused by the smaller lattice depth~[\fig{fig:Slattice}] and the reduced vertical trap depth~[\fig[b]{fig:2}] near the center of the glass cell.
For typical parameters ($a_{\text{typ}}$, $P^{\text{typ}}_G$) we measure roundtrip efficiencies for the full distance ($86\,$cm) to the science cell of >50$\%$, which is consistent with a one-way transport efficiency of 75$\%$. 
Residual deviations are most likely explained by systematic uncertainties in the atom number calibration.


To gain more insight about the dependence of the transport efficiency on the depth of the lattice [\fig[c]{fig:3}] and the acceleration  [\fig[d]{fig:3}], we evaluate the one-way transfer efficiency in the glass cell. 
We observe a scaling with $\sqrt{P_G}$ for low efficiencies, which suggests that the transport efficiency depends linearly on the depth of the lattice potential. Moreover, with increasing acceleration it falls off approximately linearly [\fig[d]{fig:3}]. Again, this suggests a linear dependence on the lattice depth, if we consider the effective depth of the potential in the presence of a tilt that is generated by the acceleration. 
The saturation of the transport efficiency for high values of $P_G$ and low accelerations $a$ indicates that these parameters do not limit the observed transport efficiency of \SI{75}{\%}.
We conclude that this value is limited by the fraction of atoms that remain in the MOT chamber and high-temperature atoms that are lost during transport due to gravity.

\section{Generation of a Bose-Einstein condensate}

\paragraph*{Dipole trap and optical evaporation} After transporting the atoms into the glass cell the atoms are levitated with a magnetic field gradient and collected in a crossed optical dipole trap at $1064\,$nm. The beam along $y$ is elliptical with a waist of $\SI{650}{\micro m}\times\SI{80}{\micro m}$ in the $xz$-plane and a maximum power of \SI{5.7}{W}. The one along $x$ is circular with $\SI{50}{\micro m}$ waist and a maximum power of $\SI{350}{mW}$ (\fig{fig:1}).
We load the dipole trap by first ramping up the dipole trap along $y$ to full power and the Gaussian transport beam to zero in \SI{350}{ms}. 
After the transport lattice is fully removed, we ramp up the dipole beam along $x$ in \SI{100}{ms} to full power.
We then reduce the offset field to \SI{27}{G} in \SI{200}{ms}.
The end of the offset field ramp defines $t=0$ in \fig[a-c]{fig:4}. 
Next, we perform a short optical evaporation, where the Bessel transport beam is turned off (\SI{500}{ms}) and at the same time the power of the dipole beam along $y$ is reduced to \SI{2}{W}.
At this point [blue dashed line in \fig[c]{fig:4}] we have \SI{9e5}{atoms} at \SI{2}{\micro K} and a PSD of $\phi=\num{3e-3}$.
The corresponding in-situ peak density is $n_0=N\bar{\omega}^3(m\lambda_{dB}/h)^3=\SI{6e11}{/cm^3}$ \cite{hungAcceleratingEvaporativeCooling2008}, where the geometric-mean trap frequency is calibrated to be $\bar{\omega}=2\pi\times \SI{40}{Hz}$.

\begin{figure}[t!]
\includegraphics{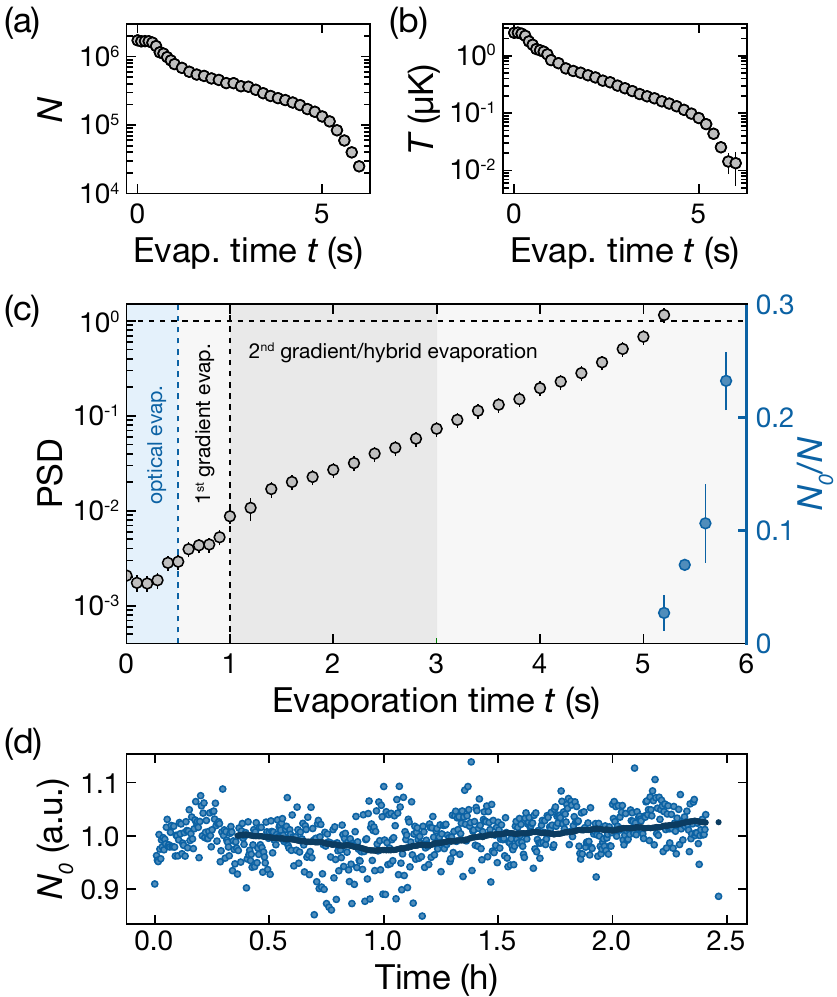}
\caption{\textbf{Evaporation to BEC}. Evolution of (a) atom number $N$, (b) temperature $T$, (c) PSD (gray dots) and BEC fraction $N_0/N$ (blue dots) during evaporation. The shaded regions indicate the evaporation steps: optical evaporation (blue) and gradient evaporation (gray). The additional optical evaporation during the second gradient evaporation is shown in dark gray. The blue dashed line shows the end of optical evaporation and the black one the end of the first gradient evaporation. 
Error bars indicate $1\sigma$-standard deviation of three repetitions.
(d) Fluctuations of the normalized BEC atom number $N_0$ over a period of $\simeq\SI{2.5}{h}$ (light blue points). The $1\sigma$-standard deviation is $<\SI{5}{\%}$. 
Dark blue points show mean values of the individual data points, averaged over 20-minute intervals. 
The $1\sigma$-standard deviation is $<\SI{2}{\%}$.}%
\label{fig:4}%
\end{figure}

\paragraph*{Gradient evaporation} To reach degeneracy we use the technique of forced evaporative cooling similar to Ref.~\cite{hungAcceleratingEvaporativeCooling2008}. We reduce the levitation gradient linearly in two successive steps to tilt the dipole potential. This allows us to keep large trapping frequencies and hence large thermalization rates during evaporation. In \fig[c]{fig:4} we show the PSD during evaporation, which was evaluated using the calibrated atom numbers [\fig[a]{fig:4}], trap frequencies and temperatures [\fig[b]{fig:4}]. 
We optimize the parameters of the evaporation ramps experimentally by maximizing the evaporation efficiency $\eta=-\ln(\phi_f/\phi_i)/\ln(N_f/N_i)$ after each step; here the indices $i (f)$ denote the parameters at the beginning (end) of the evaporation step. We find the following optimized sequence (Fig.~\ref{fig:S6} in~\cite{supplement}): During the first step, the gradient is reduced linearly to \SI{11.5}{G/cm} in \SI{500}{ms}. The second step is a hybrid evaporation scheme, where the gradient is switched off within \SI{5}{s} and the power of the dipole beam along $x$ is reduced to \SI{160}{mW} in \SI{2}{s}. The magnetic offset field is lowered to \SI{23}{G} ($a\approx$ \SI{300}{a_0}), which was found to be the optimal ratio of elastic collisions and three-body losses~\cite{weberBoseEinsteinCondensationCesium2003,chin_precision_2004}. 
This typically results in a pure BEC with \SI{2.2e4}{atoms} and evaporation efficiencies of $\eta=1.3$ for the first step and $\eta=2.5$ for the second. 

The total cycle time is \SI{11.5}{s}, comparable to other experiments without transport~\cite{hungAcceleratingEvaporativeCooling2008,kraemerEvidenceEfimovQuantum2006,grobnerNewQuantumGas2016}. We test that the duration of the evaporation ramps can be reduced further at the expense of reduced total atom numbers. With a total cycle time of \SI{7.5}{s} we obtain pure BECs with \SI{1e4}{atoms}.
Finally, to demonstrate the stability of our transport and its suitability for quantum simulation experiments, we measure the final atom number in the BEC over \SI{2.5}{h} and find that it fluctuates by $<\SI{5}{\%}$ and the mean value (averaged over 20min) drifts by $<\SI{2}{\%}$ [\fig[d]{fig:4}].
This is comparable to the atom number fluctuations before transport~\cite{supplement}, indicating that it does not induce additional instabilities.

\section{Conclusion}

In conclusion, we have demonstrated stable optical transport of heavy $^{133}$Cs atoms over a large distance of \SI{43}{cm} in less than \SI{30}{ms} with good efficiency and without observable heating. 
The transport efficiency seems to be pre-dominantely limited by the temperature of the atoms and potential depth along the vertical direction. Larger accelerations may further require the implementation of smoother frequency ramps. 
The fast transport setup demonstrated here enables short cycle times, which will be beneficial for improved statistics in future experiments. The cycle time could be reduced further by implementing additional all-optical cooling techniques~\cite{hu_creation_2017,urvoy_direct_2019,solano_strongly_2019} to reduce the loading and evaporation times in the dipole traps.
Our design further facilitates large optical access, enabling the installation of high-NA objectives for single-atom single-site resolved imaging and manipulation of cold $^{133}$Cs atoms in optical lattices~\cite{bakr_quantum_2009, shersonSingleatomresolvedFluorescenceImaging2010, yamamotoYtterbiumQuantumGas2016, cheukQuantumGasMicroscopeFermionic2015, hallerSingleatomImagingFermions2015, edgeImagingAddressingIndividual2015, parsonsSiteResolvedImagingFermionic2015, omranMicroscopicObservationPauli2015,alberti2016}. 

\subsection*{Acknowledgements}
We acknowledge insightful discussions with Cheng~Chin and his team, Elmar~Haller and Hanns-Christoph~N\"agerl. 
The authors acknowledge Andreas Reetz for help in characterizing the Bessel-beam profile and performing calculations for the design of the setup and Jingjing Chen for help in setting up the frequency-detuning setup and characterization of the pointing stability of the Bessel beam.
This work was supported by the Deutsche Forschungsgemeinschaft 
(DFG, German Research Foundation) 
under project number 277974659 via Research Unit FOR 2414. 
The work was further funded by the Deutsche Forschungsgemeinschaft (DFG, German
Research Foundation) -- 452143229 under Germany's Excellence Strategy -- EXC-2111 -- 390814868.
T.K. was supported by the Bavarian excellence 
network ENB via the International PhD Programme of Excellence 
Exploring Quantum Matter (ExQM). 
C.R.C. acknowledges support from the ICFO-MPQ Cellex postdoctoral fellowship and from
the European Union (Marie Sk{\l}odowska-Curie–897142).
H.v.R. acknowledges support from the Hector Fellow Academy.
J.F.W. acknowledges support from the German Academic Scholarship Foundation.
C.S. has received funding from
the European Union’s Framework Programme for Research 
and Innovation Horizon 2020 (2014-2020) under the Marie Sk{\l}odowska-Curie Grant Agreement No.\,754388 (LMUResearchFellows) and from LMUexcellent,
funded by the Federal Ministry of Education and Research (BMBF) and the Free State of Bavaria under the
Excellence Strategy of the German Federal Government
and the L\"ander.


\subsection*{Data availability}
The data that support the plots within this paper and other findings of this study are available from the corresponding author upon reasonable request.

\subsection*{Code availability}
The code that supports the plots within this paper are available from the corresponding author upon reasonable request.

\bibliography{transport}

\cleardoublepage

\setcounter{figure}{0}
\setcounter{page}{1}
\setcounter{equation}{0}
\setcounter{section}{0}

\renewcommand{\thefigure}{S\the\numexpr\arabic{figure}-10\relax}
 \setcounter{figure}{10}
\renewcommand{\theequation}{S.\the\numexpr\arabic{equation}-10\relax}
 \setcounter{equation}{10}
\renewcommand{\thesection}{S\Roman{section}}

\section*{Supplemental Material}

\section{Charaterization of optical transport setup}
\label{sec:TransportSetup}
\paragraph*{Bessel beam}
To compute the vertical trap depth $V_z$ along the transport direction in \fig[a]{fig:2} in the main text, we measure the Bessel beam profile at different positions behind the axicon. To ensure that the camera is not saturated we work at low power. The axicon is made from fused silica ($n=1.45$ at \SI{1064}{nm}~\cite{malitsonInterspecimenComparisonRefractive1965}), and the incident beam has low intensity, so thermal lensing in the axicon is assumed to be negligible.

\begin{figure}
\includegraphics{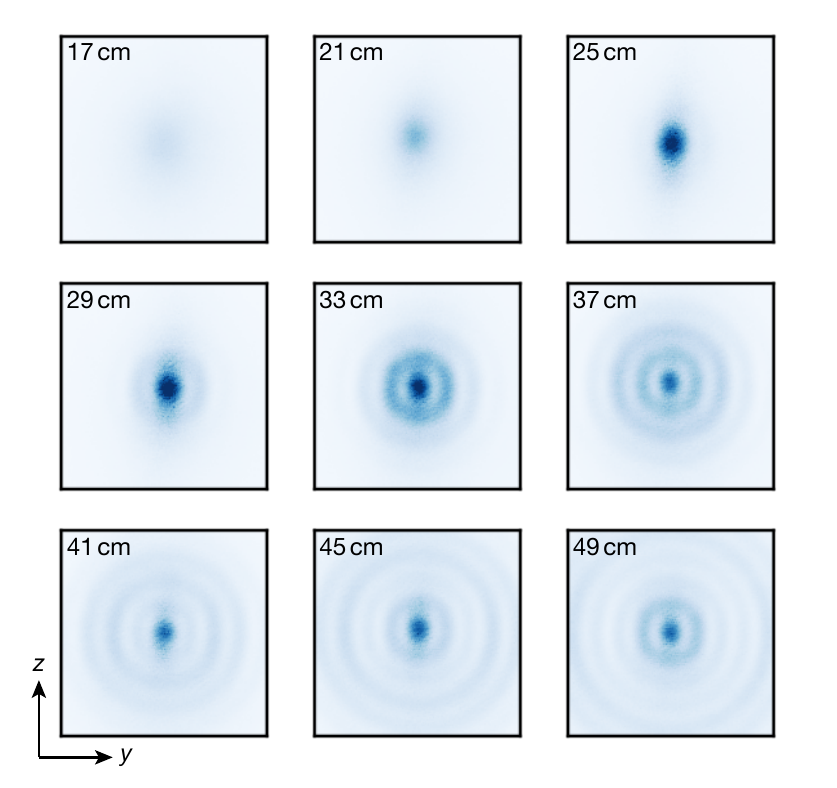}
\caption{Bessel beam profile along $x$, the transport direction, for the parameters mentioned in the main text ($\alpha=$\SI{179}{\degree}, $w_0=2.5\,$mm, $\lambda=1064\,$nm). The distance from the axicon is indicated in the top left corner of each image.}
\label{fig:S3}
\end{figure}

During the first few centimeters behind the axicon the profile of the beam is still Gaussian. About \SI{25}{cm} behind the axicon, the first diffraction ring appears and the profile becomes more Bessel-like (\fig{fig:S3}). In order to extract a peak intensity for computing the dipole trap depth, we therefore fit the images with a 2D symmetric Gaussian up to \SI{25}{cm} behind the axicon, and with a 2D Bessel function at larger distances. For the Bessel fits we use
\begin{equation}
f(y,z) = A \mathcal{J}_0\left (\sqrt{(y-y_0)^2 + (z-z_0)^2}/s_B\right )^2,
\end{equation}
where $\mathcal{J}_0$ is the zeroth-order Bessel function of the first kind and $A$, $y_0$, $z_0$ and $s_B$ are free parameters. 

From the Gaussian fits we can directly extract the waist of the beam for the computation of the depth of the dipole trap. 
For the Bessel fits, we map the fitted width to a Gaussian waist via $w_B=1.3 \sqrt{2} s_B$ [\fig[a]{fig:besselradius}]. 
We extract the relative peak intensity of the images versus distance by first subtracting a constant background from the images. 
Because the power was not kept constant for all measurements, but was increased for long distances from the axicon, we evaluate the power-normalized peak intensity via the ratio $r(x)=I_{\text{max}}/\sum_\text{px} I$, where $I_{\text{max}}$ denotes the maximum pixel value and $\sum_\text{px} I$ corresponds to the pixel sum over the whole image, after subtracting a constant background.
We then compute the relative peak intensity by normalizing $r(x)$ to the relative peak intensity after a distance of $x_\text{ref}=$\SI{13}{cm} behind the axicon, $I_\text{rel} = r/r(x_\text{ref})$.
In order to get an absolute value for the peak intensity $I_B=I_\text{rel} \cdot I_\text{ref}$ we multiply all values with the Gaussian peak intensity $I_\text{ref} = 2P/\pi w_\text{ref}^2$ at $x_\text{ref}$.
Here, the regular formula for a Gaussian beam is still applicable because the beam still has a Gaussian profile. 
The power contained in the central peak is shown in \fig[b]{fig:besselradius} for an incoming beam with $35\,$W as it was used in this work.

\begin{figure}
\includegraphics{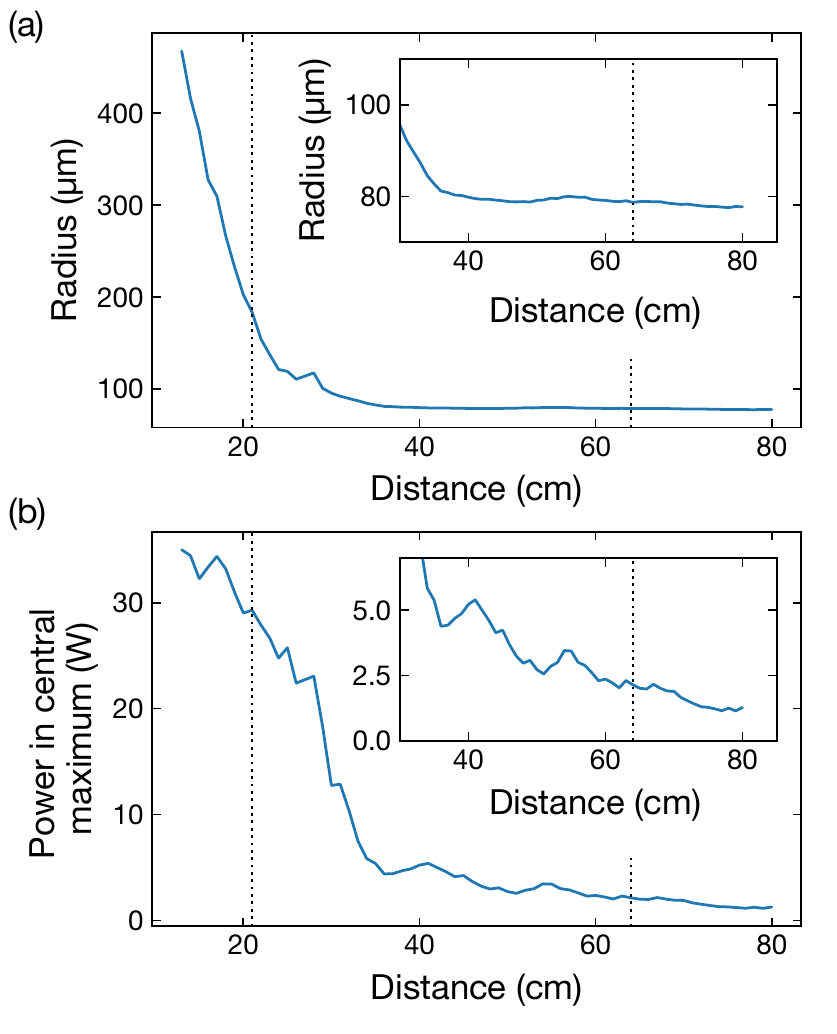}
\caption{(a) Fitted waists $w_B$ of the Bessel-shaped transport beam versus distance from the axicon. The waists are computed from the fitted radii $s_B$. The inset shows a zoom of the region, which starts at \SI{30}{cm} behind the axicon, indicating the diffraction-less propagation of the Bessel beam. (b) Power in the central maximum, extracted from the peak intensity and the fitted radii shown in (a). The input power before the axicon is \SI{35}{W} as described in the main text. The inset shows a zoom of the region starting \SI{30}{cm} behind the axicon. The dashed vertical lines in both subfigures indicate the start and end positions of the transport. }\label{fig:besselradius}
\end{figure}

\begin{figure}
\includegraphics{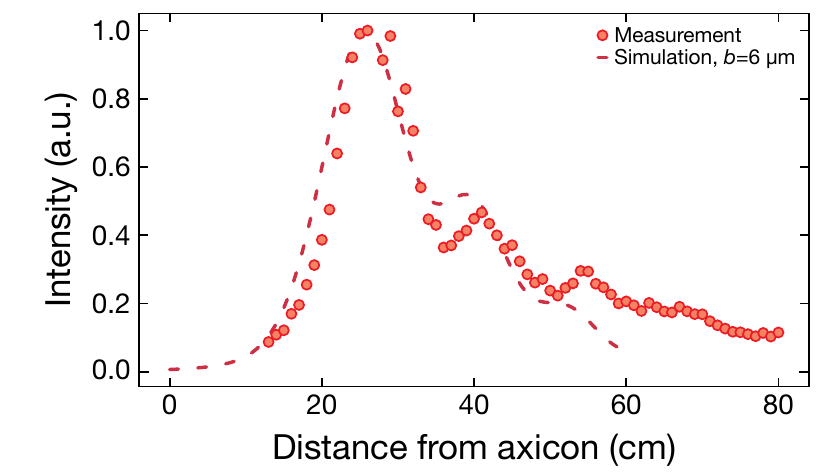}
\caption{Comparison of the measured Bessel intensity (red circles) with the theoretical intensity due to a round-tip axicon (dashed red) with $b=\SI{6}{\micro m}$.  
Both the modulation and the low intensity at short distances are due to the round tip of the axicon.}\label{fig:besselsim}
\end{figure}

An ideal axicon would result in a Gaussian intensity profile along the propagation direction. 
Imperfections can be modelled assuming a round-tip axicon~\cite{brzobohatyHighQualityQuasiBessel2008}. 
The effect of the round tip on the Bessel beam intensity is two-fold: 
First it reduces the peak power immediately behind the axicon and second it leads to a modulation of the intensity versus distance. 
Both effects are visible in our measurement.
We find reasonable agreement between theory [Ref.~\cite{brzobohatyHighQualityQuasiBessel2008}, Eq.~(6)] and experiment for a round-tip axicon with $b=\SI{6}{\micro m}$ (\fig{fig:besselsim}). 
Here $b$ is the semi major axis of the hyperbola used in approximating the round tip of the axicon.
The intensity of the Bessel beam remains low for the first few centimeters behind the axicon and we observe secondary maxima after the initial intensity peak.

\begin{figure}
\includegraphics{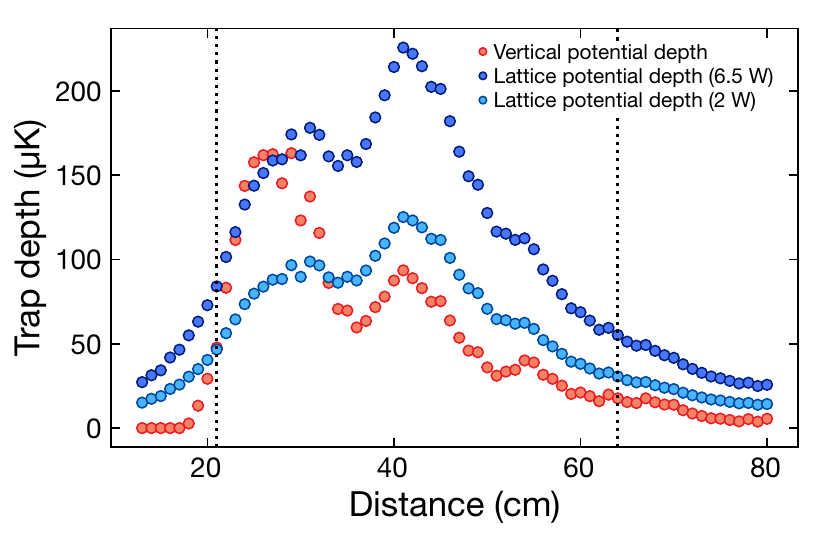}
\caption{Lattice depth for $P_G=\SI{2}{W}$ (light blue) and $P_G^\text{typ}=\SI{6.5}{W}$ (dark blue). The vertical trap depth (red) was computed for $P_G^\text{typ}$ [same data as in Fig.~\ref{fig:2}(b) in the main text]. The dashed lines indicate the start and end position of the transport.}
\label{fig:Slattice}
\end{figure}

\paragraph*{Lattice depth}
We compute the lattice depth using the measured position of the waist of the Gaussian transport beam and its power $P_G$. The radius and the power in the central peak of the Bessel beam is shown in~\fig{fig:besselradius} for a total power of the incoming beam of 35\,W.  In Fig.~\ref{fig:Slattice} we show the calculated lattice depth for two different values of the powers of the Gaussian transport beam $P_G$, together with the vertical trap depth $V_z$. The lattice depth tends to be larger than the vertical potential depth for longer transport distances.

\section{RF control of relative detuning}

\subsection{Frequency shifting setup}
For the transport lattice we use two ALS 45W Nd:YAG fiber amplifier systems. 
The two lasers are referred to as Gaussian and Bessel laser according to which of the transport beams the laser generates. 
The Gaussian laser is seeded by a low RIN NPRO Mephisto S laser from Coherent. 
A small fraction of the light from the output of the Gaussian is split off and frequency shifted using two-double pass AOMs. 
One AOM shifts the frequency by $2f_1$ within the range $f_1 \in [\num{150},175]$\,MHz the other by $2f_2$ within the range $f_2 \in [\num{200},225]$\,MHz. 
The two AOMs shift the frequency in opposite directions. 
Because the Gaussian laser is intensity stabilized using a \SI{100}{MHz} AOM, the static transport lattice is created with the two AOMs set to $2f_1=2\times \SI{175}{MHz}$ and $2f_2=2\times \SI{225}{MHz}$. 
We transport the atoms towards the glass cell by tuning the frequencies to $f_2=\SI{200}{MHz}$ and $f_1=\SI{175}{MHz}$ and transport them back to the MOT chamber with $f_2= \SI{225}{MHz}$ and $f_1=\SI{150}{MHz}$.
The output of the double DPAOM setup seeds the Bessel laser. 
This laser is not intensity stabilized beyond the internal stabilization circuitry. 

The \SI{225}{MHz} and \SI{175}{MHz} frequency is generated by two AD9914 DDS evaluations boards. 
The frequency ramps use the internal linear ramp generator of the DDS boards with a typical frequency step of \SI{550}{Hz} and a step rate of \SI{0.2}{\micro s}. 
To change the acceleration, we change the frequency step size and keep the step rate fixed. We have also tested the transport using AD9910 DDS chips from a Wieser Lab FlexDDS-NG board and found no difference in transport efficiency. 
To decrease the radio-frequency (RF) linewidth we supply an external \SI{2.5}{GHz} reference clock to the chips directly instead of relying on the chips internal PLL. 
This clock is locked to a \SI{10}{MHz} Rb reference clock, which we also use as a clock for the frequency generator (R\&S SMC100A) supplying the \SI{100}{MHz} for the Gaussian beam's intensity stabilization.
This ensures reduced relative frequency drifts between the Gaussian and the Bessel. 
Note that due to the 32 bit frequency resolution of the DDS chips, the default detuning of the Gaussian and Bessel beam is not zero but rather on the scale of a few \si{mHz}.

\subsection{Importance of RF noise}
To characterize the sensitivity of the transport to frequency noise we modulate the frequency of the \SI{100}{MHz} intensity stabilization AOM using Gaussian white noise.
The white noise is generated from an arbitrary waveform generator with a bandwidth of \SI{10}{MHz}, a peak voltage of \SI{1}{V} and a crest factor of $\approx 4.3$.
The output of the waveform generator is sent to the low frequency modulation (FM) input of the signal generator for the AOM. 
We vary the amplitude of the frequency modulation by changing the modulation bandwidth of the signal generator.
\fig[a]{fig:S4} shows the atom number measured in the MOT chamber without transport but holding the atom in the lattice for \SI{16}{ms} and the one in the glass cell after transport ($a_{\text{typ}}=\SI{2.9}{km/s^2}$, $ P^{\text{typ}}_{G}\lesssim6.5\,$W). 
We find that for modulation bandwidths $\SI{>1}{kHz}$ the atom number decreases in a similar fashion for both the atoms held in the lattice and the atoms transported.
To estimate the change in transport efficiency due to the frequency modulation, we compute the ratio of the transported and held atom number and plot it in \fig[b]{fig:S4}.
We find that the modulation also affects the transport efficiency, reducing it linearly as the modulation amplitude is increased beyond \SI{1}{kHz}.
The inset in \fig[b]{fig:S4} shows the RF linewidth versus the modulation amplitude. 
The linewidth is extracted by fitting a Lorentzian to the signal generator's output spectrum measured with a spectrum analyzer.

\begin{figure}
\includegraphics{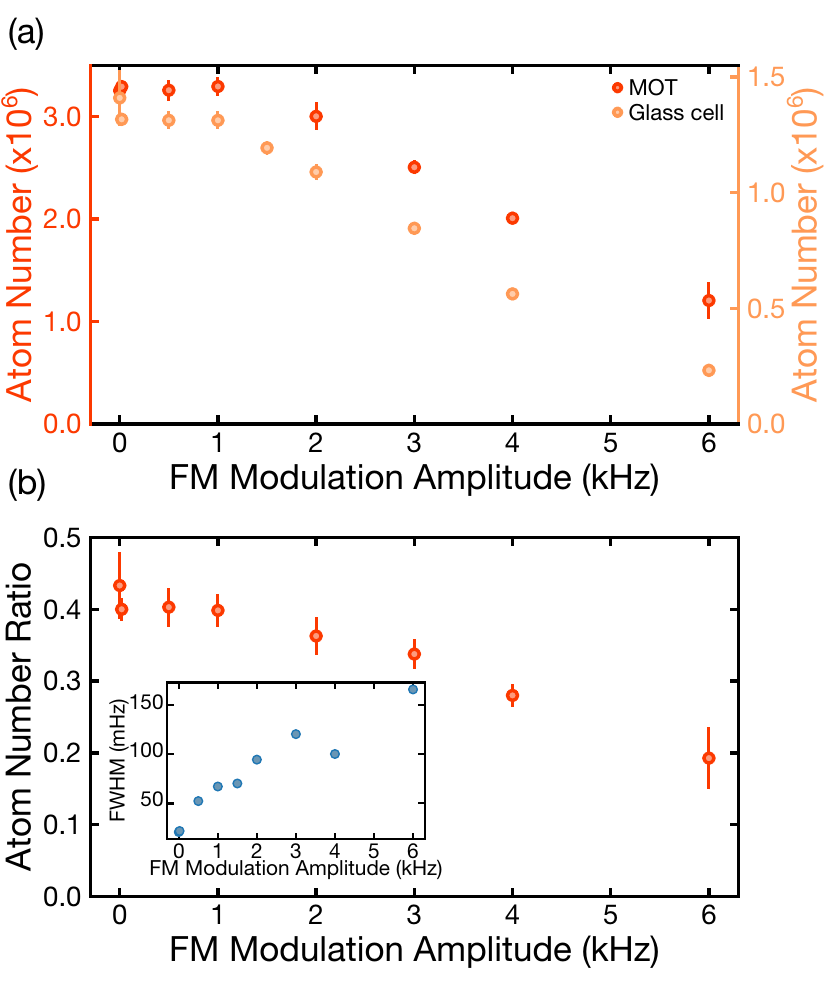}
\caption{(a) Atom number in the lattice versus frequency modulation bandwidth. The left axis shows the atoms remaining after holding them in the static lattice for \SI{25.5}{ms}, the right axis shows the atoms arriving in the glass cell after transport. (b) Ratio of the atom number in the MOT and glass cell, indicating that the transport efficiency is reduced as the noise of the lattices is increased. The inset shows the full-width-half-maximum (FWHM) linewidth of the RF source for the AOM shifting the frequency of the Gaussian beam as a function of FM modulation amplitude.}
\label{fig:S4}
\end{figure}

\section{Transport efficiency measurements}
\subsection{Extended data}
\begin{figure}
\includegraphics{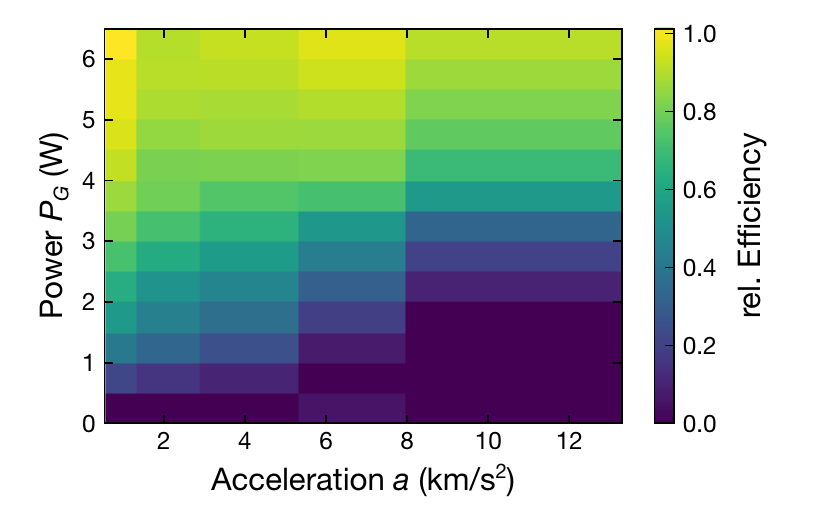}
\caption{One-way transport efficiency measurements. Figure~3(c) and (d) in the main text are crosscuts through this figure. The mean transport efficiency is normalized to the mean transport efficiency at $a=\SI{0.5}{km/s^2}$ and $P_G=\SI{6.5}{W}$. The mean is computed over five repetitions.}
\label{fig:S1}
\end{figure}

In \fig{fig:S1} we show additional measurements of the transport efficiency that were taken for different values of $P_G$ and $a$. In \fig[c]{fig:3} and (d) of the main text we show crosscuts of this 2D plot. 

\subsection{Normalization}
The measurements in \fig[c]{fig:3} and (d) of the main text are taken by measuring the atom number in the glass cell (one-way transport). 
To extract a relative efficiency, we use the mean atom number transported at the lowest acceleration of $a=\SI{0.5}{km/s^2}$ and the highest power of the Gaussian transport beam $P^\text{typ}_G=\SI{6.5}{W}$ as a reference. 
The mean is calculated using five repeated measurements. 

\begin{figure}[t!]
\includegraphics{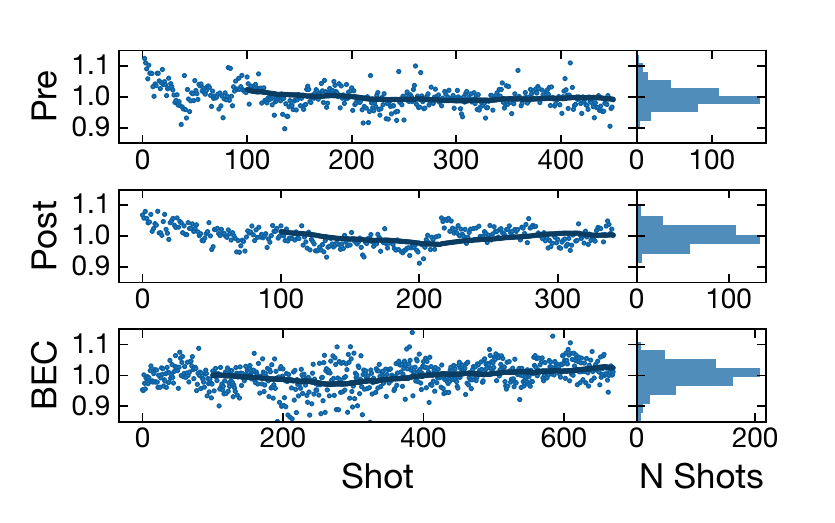}
\caption{Atom number stability before (Pre) and immediately after (Post) transport and in the BEC (same data as in Fig.~4(d) of the main text). As in the main text, the dark blue points show mean values, which were obtained by averaging the individual data points in a time window of 20\,minutes. The histograms on the left show the corresponding distributions.}
\label{fig:S5}
\end{figure}

\subsection{Guides to the eye in Fig. 3}
Before fitting the guides to the eye in \fig[c,d]{fig:3} of the main text, the data is filtered. 
For \fig[c,d]{fig:3} all points with a relative efficiency $\eta>0.75$ or $\eta<0.1$ are ignored, since we observe a saturation for low values of the acceleration and large powers, while there is essentially no transport for low values of the power $P_G$ of the Gaussian transport beam.
To account for the saturation in the acceleration data (\fig[d]{fig:3}, main text) we discard the points with accelerations smaller than $a=\SI{5}{km/s^2}$ for the measurement with a power of $P^{\text{typ}}_G=\SI{6.5}{W}$.

We fit the data in \fig[c]{fig:3} (main text) using $N(P) = b_1\sqrt{P_G} + b_0$, where $b_1$ and $b_0$ are free fit parameters, and the data in \fig[d]{fig:3} (main text) using $N(a)=\tilde{b}_1 a + \tilde{b}_0$, with free parameters $\tilde{b}_1$ and $\tilde{b}_0$.

\section{Atom number stability before and after transport}
In \fig{fig:S5} we show how the transported atom number fluctuates between repeated measurements before and directly after the transport and compare them to the fluctuations in final atom number in the BEC $N_0$. The histograms on the left show the distribution of the fluctuations for the corresponding measurements. For all data sets we have removed a slow drift, which is caused by the stabilization of the ambient temperature in the laboratory.

\section{Experimental sequence}
\label{sec:sequence}

Laser cooling of $^{133}$Cs atoms is performed on the D2 line ($|6\text{S}_{1/2} \rightarrow\text{P}_{3/2}\rangle$, $\lambda_{D2} = 852$ nm). In the following we denote the $|F=4\rangle \rightarrow |F'=5\rangle$ transition as cooling and the $|F=3\rangle \rightarrow |F'=4\rangle$ as repumping transition. 

The experimental sequence starts by loading \SI{3e7}{atoms} in a magneto-optical trap (MOT) from a Zeeman slower in \SI{3}{s}. The cooling light of the MOT is \SI{-4}{\Gamma} detuned from the cooling transition. Here $\Gamma=2 \pi\times \SI{5.2}{MHz}$ \cite{dsteckDlineData} represents the linewidth of the D2 line. Additionally, $2\,$mW of resonant light in the repumper transition are required to prevent depumping.  After MOT loading, we perform a compressed MOT stage to increase the density of the cloud. In \SI{28}{ms} we increase the magnetic field gradient from \SI{9}{G/cm} to \SI{17.5}{G/cm} and the cooler detuning to \SI{-8}{\Gamma}. Simultaneously,  we reduce the cooler and repumper power to 25\,mW and 60\,$\mu$W, respectively. We subsequently switch off the gradient and increase the cooler detuning to \SI{-22}{\Gamma} for an \SI{11.5}{ms} long molasses phase. In order to improve the cooling efficiency during this stage the residual magnetic fields are compensated with an accuracy of 50\,mG. Typically, we obtain at the end of the molasses phase \SI{3e7}{atoms}, comparable to the MOT phase, with a temperature of \SI{10}{\micro K}. 

After molasses, the atoms are optically pumped into $|F=3\rangle$, where degenerate Raman sideband cooling (dRSC) is performed in a 3D optical lattice. The Raman lattice is \SI{-20} GHz detuned from $|F=3\rangle \rightarrow |F'=2\rangle$ resonance. The lattice is generated by interfering three orthogonal beams along the $\tilde{x}$-, $\tilde{y}$- and $z$-direction. Here the ($\tilde{x}$,$\tilde{y}$) axis denote a rotation of 30$^\circ$ with respect to the coordinate system shown in \fig[a]{fig:1}. The $\tilde{y}$-lattice beam is retro-reflected after passing through a $\lambda/4$ waveplate. The beams are linearly polarized. The $\tilde{x}$ and $z$ beams are polarized such that they interfere only with the $\tilde{y}$ beam. In addition to the lattice, a circularly-polarized beam propagates along the $z$-direction. This beam is resonant with the $|F=3\rangle \rightarrow |F'=2\rangle$ transition with a power of 650\,$\mu$W. After 7\,ms of dRSC cooling we obtain \SI{2e7}{atoms} at <1\,$\mu$K, at a density of $3\times 10^{10}$\,atoms/cm$^{3}$. The following steps towards condensation are described in the main text. 

\begin{figure}
\includegraphics{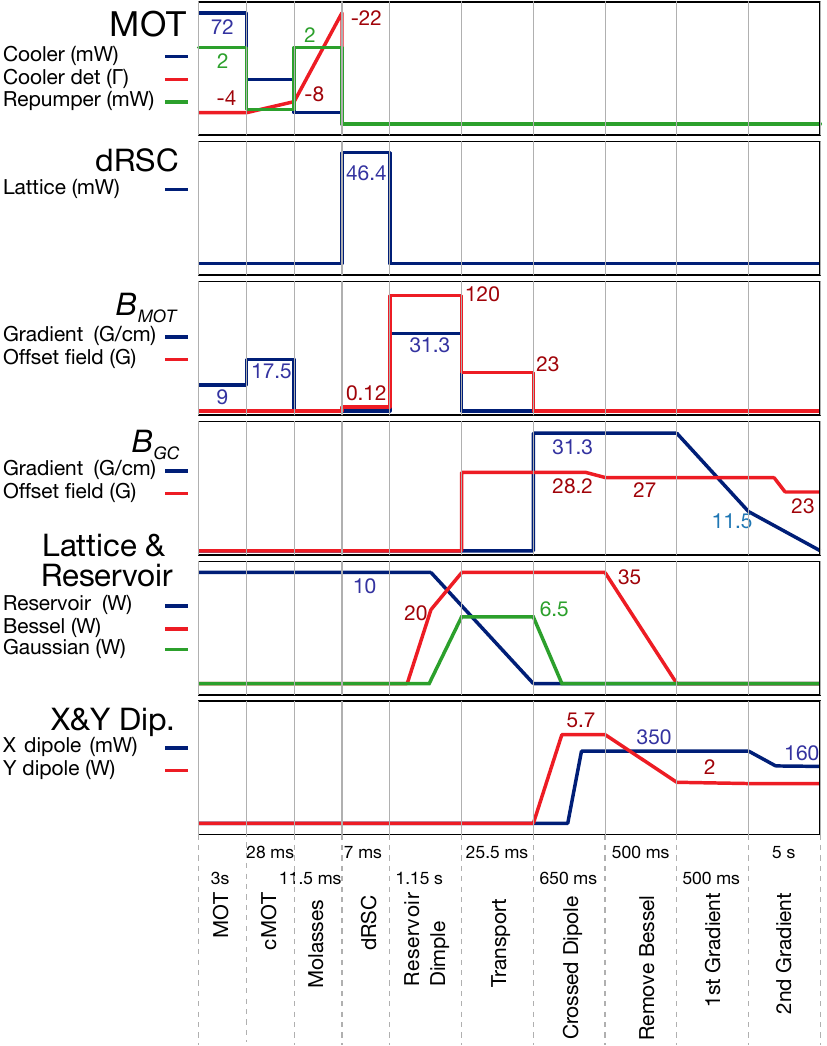}
\caption{Sketch of the sequence. The relative changes in detuning, power and magnetic field ($B$) are shown for both the MOT chamber and the glass cell (GC). The horizontal and vertical axes are not to scale in order to improve readability.}
\label{fig:S6}
\end{figure}

The detailed sequence for producing degenerate gases of cesium atoms is presented in \fig{fig:S6}. This includes the relative changes in detuning, power and magnetic fields either in the glass cell (GC) or in the MOT chamber (MOT). We sketch the respective values for the start and end of the  ramps.

\end{document}